\documentclass{aastex}
\usepackage{emulateapj5}
\usepackage{apjfonts}
\usepackage{epsf}

\slugcomment{UTAP-437}
\shorttitle{ARRIVAL DISTRIBUTION OF UHECRS}
\shortauthors{YOSHIGUCHI ET AL.}
\begin{document}
\title{Arrival Distribution of
Ultra-High Energy Cosmic Rays:\\
Prospects for the Future}
%
\author{Hiroyuki Yoshiguchi\altaffilmark{1}, Shigehiro
Nagataki\altaffilmark{1,2}, and Katsuhiko Sato\altaffilmark{1,2}}

\altaffiltext{1}{Department of Physics, School of Science, the University
of Tokyo, 7-3-1 Hongo, Bunkyoku, Tokyo 113-0033, Japan}
\altaffiltext{2}{Research Center for the Early Universe, School of
Science, the University of Tokyo,
7-3-1 Hongo, Bunkyoku, Tokyo 113-0033, Japan}

\email{hiroyuki@utap.phys.s.u-tokyo.ac.jp}
%
\received{}
\accepted{}
\begin{abstract}
We predict the arrival distribution of UHECRs above $4 \times 10^{19}$
eV with the event number
expected by future experiments in the next few years.
We perform event simulations with the source model which is adopted in
our recent study and can explain the current AGASA observation.
At first, we calculate the harmonic amplitude and the two point
correlation function for the simulated event sets.
We find that significant anisotropy on large angle scale will be
observed when $\sim 10^3$ cosmic rays above $4 \times 10^{19}$ eV
are detected by future experiments.
The Auger array will detect cosmic rays with this event number in
a few years after its operation.
The statistics of the two point correlation function will also
increase.
The angle scale at which the events have strong
correlation with each other corresponds to deflection angle of UHECR
in propagating in the EGMF, which in turn can be determined by the
future observations.
We further investigate the relation between the number of events
clustered at a direction and the distance of their sources.
Despite the limited amount of data, we find that the C2 triplet events
observed by the AGASA may originate from the source within 100 Mpc
from us at $2 \sigma$ confidence level.
Merger galaxy Arp 299 (NGC 3690 $+$ IC 694) is the best candidate
for their source.
If data accumulate,
the UHECR sources within $\sim 100$ Mpc can be identified
from observed event clusterings significantly.
This will provide some kinds of information about poorly known
parameters which influence the propagation of UHECRs, such as
extragalactic and galactic magnetic field, chemical composition of
observed cosmic rays.
Also, we will reveal their origin with our
method to identify the sources of UHECR.
Finally, we predict the arrival distribution of UHECRs above
$10^{20}$ eV, which is expected to be observed if the current HiRes
spectrum is correct, and discuss their statistical features and
implications.
\end{abstract} 
%
\section{INTRODUCTION} \label{intro}
The AGASA observation of ultra high energy cosmic rays (UHECRs) above
$10^{19}$ eV reveals at least two features.
The cosmic-ray energy spectrum does not show the GZK cutoff
\citep*{greisen66,zatsepin66} because of
photopion production with the photons of the cosmic microwave
background (CMB) and extends above $10^{20}$ eV
\citep*{takeda98}.
On the other hand, their arrival distribution seems to be isotropic
on a large scale with a statistically significant small scale
clustering \citep*{takeda99}.
The current AGASA data set of 57 events above 4 $\times 10^{19}$ eV
contains four doublets and one triplet within
a separation angle of 2.5$^\circ$.
Chance probability to observe such clusters under an isotropic
distribution is only about 1 $\%$ \citep*{hayashida00}.

Recently, the High Resolution Fly's Eye \citep*[HiRes;
][]{wilkinson99} reports the cosmic ray flux with the GZK
cut-off around $10^{20}$ eV \citep*{abu02}.
At present, it is very difficult to draw a conclusion about the
existence or non-existence of the GZK cutoff, because both the two
experiments have detected only a handful of events above this energy.
On the other hand, there are new large-aperture detectors under
development, such as South and North Auger project \citep*{capelle98},
the EUSO \citep*{euso92}, and the OWL \citep*{owl00} experiments.
The detection or non-detection of the GZK cutoff in the cosmic-ray
spectrum remains open to investigation by these future generation
experiments.

Potential models of UHECR origin are constrained by their ability to
reproduce the measured energy spectrum and the arrival
distribution observed by the AGASA.
In our recent work \citep*[][hereafter Paper I]{yoshiguchi02a}, we
perform numerical simulations for propagation of UHE protons in
intergalactic space, and examine whether the present AGASA observation
can be explained by a bottom-up scenario in which the source
distribution of UHECRs is proportional to that of galaxies.
We use the Optical Redshift Survey
\citep*[ORS; ][]{santiago95} to construct realistic source models of
UHECRs.
We can construct realistic source models of UHECRs by using the
galaxy sample,
because the astrophysical candidates of UHECR sources, such as active
galactic nuclei \citep*[AGN, ][]{halzen97}, gamma-ray bursts
\citep*[GRB, ][]{waxman95,waxman00}, and colliding galaxies
\citep*{cesarsky92,smi02} are in connection with the galaxies.
For example, AGNs are considered as supermassive black holes in the
centers of the galaxies.
A GRB occurs in a galaxy, and so on.

In Paper I, we calculate both the energy spectrum and arrival
directions of UHE protons, and compare the results with the AGASA
observation.
We find that the arrival distribution of UHECRs become to be most
isotropic as restricting sources to luminous galaxies $(M_{\rm
{lim}}=-20.5)$.
This is because luminous galaxies in the Local Super Cluster (LSC)
distribute outward than faint galaxies, contrary to general clusters
of galaxies \citep*{yoshiguchi02b}.
However, it is not isotropic enough to be consistent with the
AGASA observation, even for $M_{\rm {lim}}=-20.5$.
In order to obtain sufficiently isotropic arrival distribution,
we randomly select sources, which contribute to the observed cosmic
ray flux, from the ORS sample more luminous than $-20.5$ mag, and find
that the isotropic arrival distribution of UHECRs can be reproduced in
the case that the number fraction of $\sim 1/50$ of the sample is
selected as UHECR sources.
In terms of the source number density, this constraint
corresponds to $\sim 10^{-6}$ Mpc$^{-3}$.

We further find that the small scale anisotropy can not be well
reproduced in the case of strong extragalactic magnetic field ($B
\ge 10$ nG).
This is because the correlation at small scale between events which
originate from a single source is eliminated, or the correlation continues
to larger angle scale, due to large deflection when propagating in the
EGMF from sources to the earth.

However, we should comment on the studies by other workers.
\cite{isola02} and \cite{sigl03} study the propagation of UHE protons
in strong EGMF ($\sim 1 \mu$ G) in the LSC, assuming local enhancement
of UHECR sources in the LSC.
The strong EGMF of $\sim 1 \mu$ G leads to substantial deflections
of UHECRs, which are better for explaining the observed isotropic
distribution of UHECRs.
However, the consistency of small-scale anisotropy and also
large-scale isotropy predicted by their scenarios with the AGASA
observation is marginal and somewhat worse than that predicted by our
scenario in Paper I.
There are also suggestions that the observed energy spectra have the
imprints of UHE proton interaction with the CMB photons as beginning
of the GZK cutoff at $E \sim 4 \times 10^{19}$ eV
\citep*{berezinsky02b,berezinsky03,marco03}.
In the presence of local enhancement of UHECR sources in the LSC, this
feature of the observed spectra seems to be difficult to be
reproduced.
The AGASA observation may imply that UHECRs propagate along nearly
straight lines in intergalactic space.
However we can not draw any firm conclusion because of the limited
amount of data.
The next generation experiments will clarifies the situation with a
large amount of data.

If local enhancement of UHECR sources in the LSC
\citep*{sigl99,lemoine99,isola02,sigl02,sigl03} is disfavored from
the observations, there is no way that explains the observed extension
of the cosmic-ray spectrum beyond the GZK cutoff.
Our conclusion in Paper I is that a large fraction of cosmic rays above
$10^{20}$ eV observed by the AGASA experiment might originate in the
top-down scenarios, or that the energy spectrum measured by the Hires
experiment might be better.

As mentioned above, many future new experiments are under development.
These experiments will provide us a large number of data, at least below
$10^{20}$ eV, and allow us to discuss the features of the arrival
distribution of UHECRs and determine which model of the UHECR origin
can explain the observations with better statistical significance.
In this paper, we predict the arrival distribution of UHECRs above
$4 \times 10^{19}$ eV with the event number expected by future
experiments in the next few years.
We perform event simulations with the source model which can explain
the current AGASA observation.
It is noted that our prediction is not the exact arrival directions of
each UHECR but the statistical features of the arrival distribution,
because there are degrees of freedom of randomly selecting the UHECR
sources from the ORS sample.
At first, we examine how much the future experiments decrease the
statistical uncertainty of the cosmic-ray spectrum at the highest
($\sim 10^{20}$ eV) energies \citep*{marco03}.
Next, we calculate the harmonic amplitude and the two point
correlation function, and demonstrate that observational constraints
on the model of UHECR origin become severer by new experiments.
We further investigate the relation between the number of events
clustered at a direction and the distance of their source.
Such analysis has never performed before.
Implications of the results are discussed in detail.
Finally, we also predict the arrival distribution of UHECRs above
$10^{20}$ eV, which is expected to be observed if the current HiRes
spectrum is correct, and discuss their statistical features and
implications.

In section~\ref{model}, we describe our method of calculation. 
Results are shown in section~\ref{result}.
In section~\ref{summary}, we summarize the main results and discuss
their implications.


\section{NUMERICAL METHOD} \label{model}
\subsection{Numerical Simulation}
\label{sim}

This subsection provides the method of Monte Carlo simulations
for propagation of UHE protons in intergalactic space.
We use the same numerical approach used in Paper I.
Detailed explanations are presented in Paper I.

UHE protons below $\sim 8 \times 10^{19}$ eV lose their energies
mainly by pair creations and above it by photopion production
\citep*{berezinsky88,yoshida93} in collisions with photons of the CMB.
The pair production can be treated as a continuous loss process
considering its small inelasticity ($\sim 10^{-3}$).
We adopt the analytical fit functions given by \cite{chodorowski92}
to calculate the energy loss rate for the pair production
on isotropic photons.
On the other hand, protons lose a large fraction of their energy in
the photopion production.
For this reason, its treatment is very important.
We use the interaction length and the energy distribution of
final protons as a function of initial proton energy
which is calculated by simulating the photopion
production with the event generator SOPHIA \citep*{sophia00}.

The EGMF are little known theoretically and observationally.
There is the upper limit for the strength and correlation length
of the universal EGMF, $B \cdot l_{\rm c}^{1/2} < 1 {\rm nG} (1{\rm
Mpc})^{1/2}$, as measured by Faraday rotation of radio signals from
distant quasars \citep*{kron94}.
However, simple analytical arguments based on magnetic flux freezing,
and large scale structure simulations passively including the magnetic
field \citep*{kulsrud97} demonstrate that the magnetic field is most
likely as structured as are the baryons.
The local EGMF as strong as $\sim 1 \mu$G in sheets and filaments of large
scale galaxy distribution, such as in the LSC, are compatible with existing
upper limits on Faraday rotation \citep*{ryu98,blasi99}.
It is suspected that the arrival distribution of UHECRs depends on the
fields in the immediate environment of the observer.

However, we may expect that the effects of such strong EGMF on UHECR
arrival directions may be small in our source scenario.
In the source model by \cite{isola02} and \cite{sigl03}, all sources
are assumed to be in the LSC.
Thus, in their model the effect of the strong EGMF on the deflection
angles of UHECRs is reported to be large.
On the contrary, we show in Paper I that the source number density
$\sim 10^{-6}$ Mpc$^{-3}$ is favored in order to explain the arrival
distribution of UHECRs observed by the AGASA.
In this case, there is no source in the LSC.
Accordingly, almost all of the paths of the observed UHECRs from
sources to the earth are not in the LSC but outside it in our model,
because the structure of the strong EGMF in the LSC would be like
two-dimensional sheet of large scale galaxy distribution, as shown in
\cite{sigl03}.
We also note that Faraday rotation measurement gives not the absolute
value of the strength of the EGMF, but only its upper limit.

We further find that small scale clustering can not be well
reproduced in the case of strong EGMF $(B > 10{\rm nG})$.
If the local strong EGMF affects the arrival directions of UHECRs,
small scale clustering observed by the AGASA may not be obtained.
Thus, we assume that the local strong EGMF in the LSC, even if it
exists with the structure like two-dimensional sheet, does not affect
the arrival directions of UHECRs, and adopt a homogeneous random
turbulent magnetic field with $(B,l_{\rm c})=(1 {\rm nG}, 1 {\rm Mpc})$.

Of course, we can not draw conclusion on the effects of the strong
EGMF, considering the limited amount of observed data.
However, since the model of local enhancement of UHECR sources
\citep*{isola02,sigl03} predicts the emergence of large-scale
anisotropy which reflects the spatial structure of the LSC, we will be
able to draw firm conclusion by comparing the prediction of our source
scenario presented in this paper with the results of the future
experiments.
This is one of the purposes of the present work.

We also note that numerical simulations of UHECR propagation in
inhomogeneous EGMF over cosmological distances are highly
time-consuming.
(On the other hand, \cite{isola02} and \cite{sigl03} perform numerical
simulations of UHECR propagation over $\sim$ 50 Mpc.)
With the assumption of homogeneous EGMF, we can perform numerical
simulation of UHECR propagation under spherical symmetry.
Provided that the EGMF is inhomogeneous, however, the propagation of
UHECRs from a single point source has no longer spherical symmetry.
As a result, we have to specify the earth position in the universe.
We also have to choose the detector (earth) size so small enough for
us to accurately calculate the arrival directions.
In this case, the number fraction of UHECRs arriving at the earth to
injected ones is extremely small.
This requires the number of particle to be propagated several orders
of magnitudes higher than that used in this study, which takes
enormous CPU time.
Detailed study on the effects of the strong EGMF in the LSC are beyond
the scope of this paper and left for future study.

We assume turbulent magnetic field with power-law
spectrum $\langle B^2(k)\rangle \propto k^{n_H}$ for
$2 \pi / l_{\rm c} \le  k \le 2 \pi / l_{\mbox{{\scriptsize cut}}}$
and $\langle B^2(k)\rangle =0$ otherwise,
where $l_{\mbox{{\scriptsize cut}}}$ characterizes
the numerical cut-off scale.
We use $n_H=-11/3$ corresponding to the Kolmogorov spectrum.
Physically one expect $l_{\mbox{{\scriptsize cut}}} \ll
l_{\rm c}$, but we set $l_{\mbox{{\scriptsize cut}}} = 1/8 \times l_{\rm c}$
in order to save the CPU time.
The universe is covered with cubes of side $l_{\rm c}$.
For each of the cubes,
Fourier components of the EGMF are dialed on a cubic cell
in wave number space, whose side is $2 \pi / l_{\rm c}$,
with random phases according to the Kolmogorov spectrum,
and then Fourier transformed onto the corresponding cubic cell in real space.
We create the EGMF of 20 $\times$ 20 $\times$ 20 cubes of side $l_{\rm c}$,
and outside it,
adopt the periodic boundary condition
in order to reduce storage data for magnetic field components.
Similar methods for the turbulent magnetic fields have been
adopted \citep*{sigl99,lemoine99,isola02}.
In this study, we neglect the effects of the galactic magnetic field.
We will conduct studies on its effects in forthcoming paper.

Finally, we explain how the energy spectrum and the arrival directions
of UHECRs are calculated.
At first, protons with a flat energy spectrum are injected
isotropically at a given point within the range of ($10^{19.5}$ -
$10^{22}$)eV.
5000 protons are injected in each of 26 energy bins, that is,
10 bins per decade of energy.
Then, UHE protons are propagated in the EGMF over $1$ Gpc for $15$ Gyr.
Weighted with a factor corresponding to a $E^{-2}$ power law spectrum,
this provides distribution of energy, deflection angle, and time delay
of UHECRs as a function of the distance from the initial point.
In this paper, we use the distribution of energies and deflection angles
integrated over the time delay, assuming that the cosmic ray flux
at the earth is stationary.
With this distribution, we can calculate the energy spectrum and the
arrival directions of UHECRs injected at a single UHECR source.
Then, summing contributions from all the sources (see the
section~\ref{source}), we obtain the angular probability distributions
of UHECRs as a function of their energies.
According to this angular probability distributions, we simulate the
energy spectrum and the arrival directions of UHECRs.

\subsection{Source Distribution}
\label{source}

In this study, we assume that the source distribution of UHECRs
is proportional to that of the galaxies.
We use the realistic data from the ORS \citep*{santiago95} galaxy
catalog.
As mentioned in section~\ref{intro}, we show in Paper I that the
arrival distribution of UHECRs observed by the AGASA can be reproduced
in the case that the number fraction of $\sim 1/50$ of the ORS
galaxies more luminous than $M_{\rm lim}=-20.5$ is selected as UHECR
sources.
We consider only the prediction of this source model throughout the
paper.
It is unknown how much an ultimate UHECR source
contribute to the observed cosmic ray flux.
In paper I, we thus consider the two cases in which all galaxies are
the same, and they inject cosmic rays proportional to their absolute
luminosity.
However, we find that the results in the two cases do not differ from
each other, as far as we focus on the luminous galaxies as UHECR
sources.
Accordingly, we restrict ourselves to the case that all galaxies
inject cosmic rays with the same amount.

In order to calculate the energy spectrum and the distribution of arrival
directions of UHECRs realistically, there are two key elements
of the galaxy sample to be corrected.
First, galaxies in a given magnitude-limited sample are biased tracers
of matter distribution because of the flux limit.
Although the sample of galaxies more luminous than $-20.5$ mag is
complete within 80 $h^{-1}$ Mpc, it does not contain galaxies
outside it for the reason of the selection effect, where $h$ is the
Hubble constant divided by 100 km s$^{-1}$ and we use $h=0.75$.
We distribute sources of UHECRs outside 80 $h^{-1}$ Mpc homogeneously,
and calculate their amount from the number of galaxies inside it.
Second, our ORS sample does not include galaxies
in the zone of avoidance ($|b|<20^{\circ}$).
In the same way, we distribute UHECR sources in this region homogeneously,
and calculate its number density from the number of galaxies in the
observed region.

\subsection{Statistical Methods}\label{statistics}

In this subsection, we explain the three statistical quantities,
the harmonics analysis for large scale anisotropy \citep*{hayashida99},
the two point correlation function for small scale anisotropy, and the
correlation value for investigation of the correlation between the
events and their sources defined in our previous study
\citep*{ide01}.

The harmonic analysis to the right ascension distribution of events
is the conventional method to search for global anisotropy
of cosmic ray arrival distribution.
For a ground-based detector like the AGASA and the Auger, the almost
uniform observation in right ascension is expected.
The $m$-th harmonic amplitude $r$ is determined by fitting the distribution
to a sine wave with period $2 \pi /m$.
For a sample of $n$ measurements of phase,
$\phi_1$, $\phi_2$, $\cdot \cdot \cdot$, $\phi_n$
(0 $\le \phi_i \le 2 \pi$), it is expressed as
\begin{equation}
r = (a^2 + b^2)^{1/2}
\label{eqn121}
\end{equation}
where, $a = \frac{2}{n} \Sigma_{i = 1}^{n} \cos m \phi_i  $,
$b = \frac{2}{n} \Sigma_{i = 1}^{n} \sin m \phi_i  $.
We calculate the harmonic amplitude for $m=1-4$ from a set of events
generated according to predicted probability density distribution
of arrival directions of UHECRs.

If events with
total number $n$ are uniformly distributed in right ascension, the chance
probability of observing the amplitude $\ge r$ is given by,
\begin{equation}
P = \exp (-k),
\label{eqn13}
\end{equation}
where
\begin{equation}
k = n r^2/4.
\label{eqn14}
\end{equation}
The current AGASA 57 events is consistent with isotropic source
distribution within 90 $\%$ confidence level
\citep*{takeda99,hayashida00}.
We therefore compare the harmonic amplitude for $P = 0.1$ with the
model prediction, and estimate the event number at which large scale
anisotropy of the UHECR arrival distribution become significant.

The two point correlation function $N(\theta)$ contains
information on the small scale anisotropy.
We start from a set of generated events or the actual AGASA data.
For each event, we divide the sphere into concentric bins of
angular size $\Delta \theta$, and count the number of events falling
into each bin.
We then divide it by the solid angle of the corresponding bin,
that is,
\begin{eqnarray}
N ( \theta ) = \frac{1}{2 \pi | \cos \theta  - \cos (\theta + \Delta \theta)
|} \sum_{ \theta
\le  \phi \le \theta + \Delta \theta }  1 \;\;\; [ \rm  sr ^{-1} ],
\label{eqn100}
\end{eqnarray}
where $\phi$ denotes the separation angle of the two events.
$\Delta \theta$ is taken to be $1^{\circ}$ in this analysis.
The AGASA data shows strong correlation at small angle $(\sim 2^{\circ})$
with 5 $\sigma$ significance of deviation from an isotropic distribution
\citep*{takeda99,hayashida00}.

We use the correlation value, defined in our previous study
\citep*{ide01}, in order to investigate statistically the
similarity between the arrival distribution of UHECRs and the source
distribution.
The correlation value, $\mit\Xi$, between two distributions $f_{\rm
e}$ and $f_{\rm s}$, is defined as
\begin{equation}
\mit\Xi (f_{\rm e},f_{\rm s}) \equiv
\frac{\rho (f_{\rm e},f_{\rm s})}{\sqrt{\rho (f_{\rm e},f_{\rm e})
\rho (f_{\rm s},f_{\rm s})}},
\label{def_cor}
\end{equation}
where
\begin{equation}
\rho (f_a,f_b) \equiv \sum _{j,k}\left( \frac{f_a(j,k)-\bar f_a}{\bar f_a }
\right)\left( \frac{f_b(j,k)-\bar f_b }{\bar f_b } \right)  \frac{\Delta
\Omega (j,k)}{4\pi}.
\label{def_rho} 
\end{equation}
Here subscripts $j$ and $k$ discriminate each cell of the sky,
$\Delta\Omega (j,k)$ denotes the solid angle of the ($j,k$) cell, and
$\bar f$ means the average of $f$.
In equation~\ref{def_cor}, $f_{\rm e}$ and $f_{\rm s}$ represent the
distribution of the simulated events and the sources, respectively.
In this study, the size of the cell
is chosen to be $1^{\circ} \times 1^{\circ}$.
The meaning of $\mit\Xi$ is as follows. By definition, $\mit\Xi$ ranges
from $-1$ to $+1$. When $\mit\Xi =$ +1(-1), two distributions
are exactly same (opposite).
When $\mit\Xi = $ 0, we can not find any resemblance between two
distributions.

\section{RESULTS} \label{result}

\subsection{Statistical Significance of the Energy Spectrum at $\sim
10^{20}$ eV}\label{spectrum}

Before we discuss the future prospect of the UHECR arrival
distribution, we examine how much the future experiments decrease the
statistical uncertainty of the cosmic-ray spectrum at the highest
energy range $(\sim 10^{20} \; {\rm eV})$.
In figure~\ref{esp_now}, we show the energy spectrum predicted by a
specific source scenario in the case
that the number fraction of $1/50$ of the ORS galaxies more luminous
than $M_{\rm{lim}}=-20.5$ is selected as UHECR sources.
The injection spectrum is taken to be $E^{-2}$.

\vspace{0.5cm}
\centerline{{\vbox{\epsfxsize=8.0cm\epsfbox{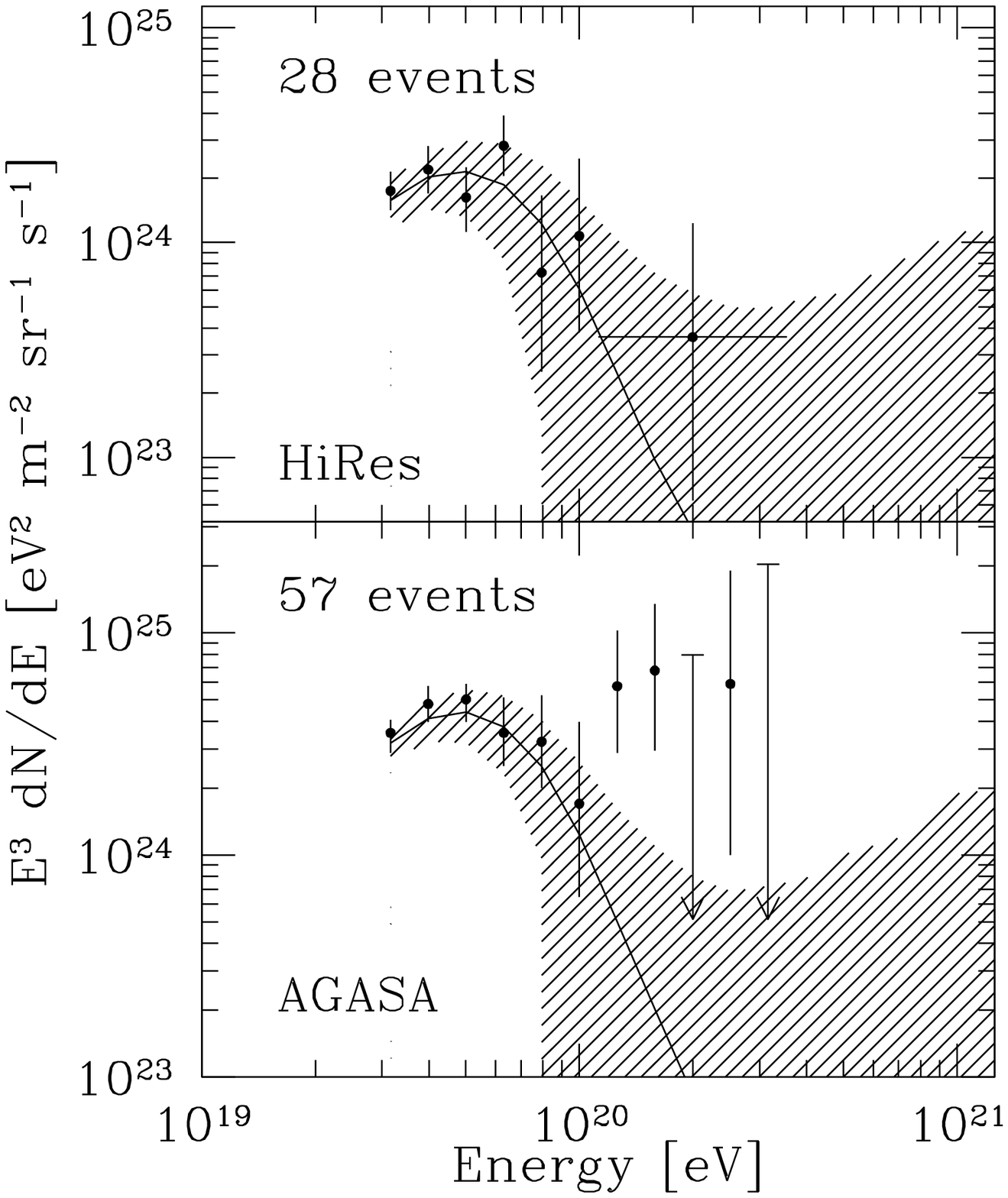}}}}
\figcaption{
Energy spectrum with injection spectrum $E^{-2}$,
predicted by a specific source scenario in the case
that the number fraction of $1/50$ of the ORS galaxies more luminous
than $M_{\rm{lim}}=-20.5$ is selected as UHECR sources.
The simulations are performed with the fixed event numbers above
$4 \times 10^{19}$ eV.
The shaded region indicates the statistical error due to the finite
number of simulated events.
We also show the observed cosmic-ray spectrum by the Hires
\citep*{abu02} and the AGASA \citep*{hayashida00}.
\label{esp_now}}
\vspace{0.5cm}

\vspace{0.5cm}
\centerline{{\vbox{\epsfxsize=8.0cm\epsfbox{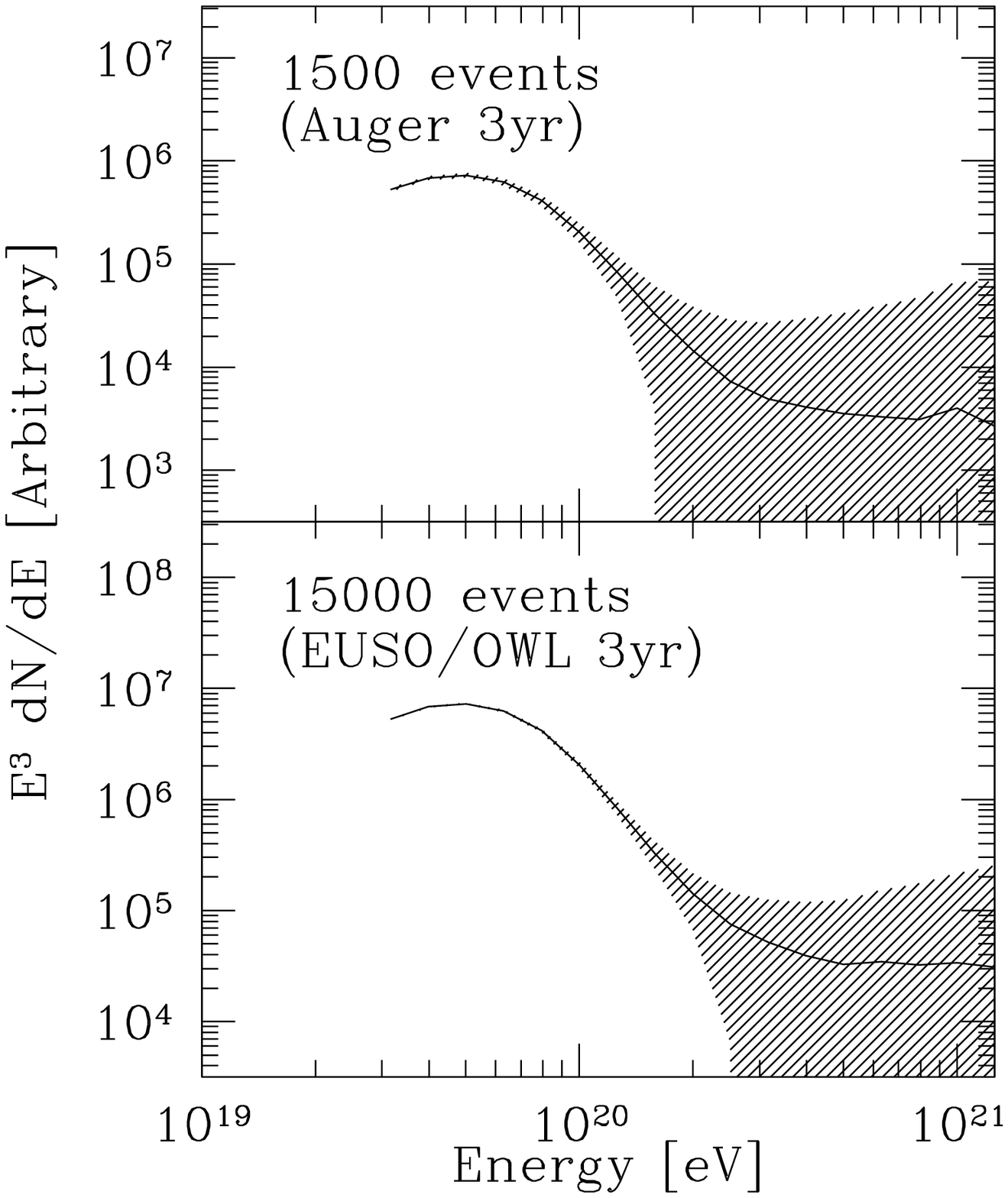}}}}
\figcaption{
Same as figure~\ref{esp_now}, but with the total number of events
expected by 3 year operation of the Auger and the EUSO/OWL.
\label{esp_future}}
\vspace{0.5cm}

\cite{berezinsky02} show that predicted flux of UHECRs
fall short of the observed flux below $10^{19.5}$ eV in the case of
injection spectrum $E^{-2}$.
However, there may be UHECR production sites in which
maximum energy of cosmic ray achieved is lower than $10^{19.5}$ eV.
These components may substantially contribute to the cosmic ray flux
below $10^{19.5}$ eV.
Throughout the paper, we assume that these components
bridge the gap between the observed flux
and the predicted one with the injection spectrum $E^{-2}$,
and restrict ourselves to cosmic rays only above $10^{19.5}$ eV.

In figure~\ref{esp_now}, event simulations are performed with the fixed
event numbers above $4 \times 10^{19}$ eV.
Typically, we perform 10000 such simulations.
The shaded region indicates the statistical error due to the finite
number of simulated events.
The spectrum measured by the HiRes is consistent with our model
prediction, while that of the AGASA is not.
However, the statistical significance of deviation from the prediction
of our source scenario is about only $\sim 2 \sigma$.
The region of the energy spectrum dominated by
statistical fluctuation is moved to higher energies with increasing
the data, as shown in figure~\ref{esp_future}.
It is noted that the future experiment, such as the Auger and the
EUSO/OWL, would detect $\sim 500$ and $\sim 5000$ events above $4
\times 10^{19}$ eV per year, respectively.
This high statistics will allow us to conclude the presence or
absence of the GZK cutoff in the cosmic-ray spectrum in the next
few years.
Similar conclusion is obtained in \cite{marco03}.

\subsection{Arrival Distribution of UHECRs above 4$\times 10^{19}$eV}
\label{arrival}

\begin{figure*}
\begin{center}
\epsscale{1.4} 
\plotone{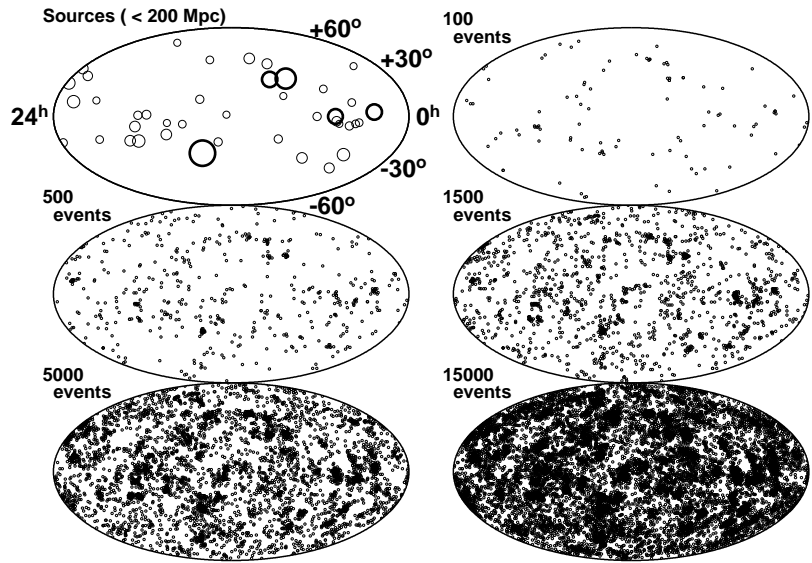} 
\caption{
Realizations of arrival directions of UHECRs above $4 \times 10^{19}$
eV predicted by
a specific source scenario in the case that the number fraction of
$1/50$ of the ORS galaxies more luminous than $M_{\rm{lim}}=-20.5$ is
selected as UHECR sources.
Distribution of selected sources within $200$ Mpc is also shown as
circles of radius inversely proportional to their distances.
Only the sources within $100$ Mpc are shown with bold lines.
\label{event}}
\end{center}
\end{figure*}

In this subsection, we present the results of the numerical calculations
for the arrival distributions of UHECRs.
Figure~\ref{event} shows realizations of the UHECR arrival direction
above $4 \times 10^{19}$ eV predicted by a specific source scenario in
the case that the number fraction of $1/50$ of the ORS galaxies more
luminous than $M_{\rm{lim}}=-20.5$ is randomly selected as UHECR
sources.
Distribution of selected sources within $200$ Mpc is also shown as
circles of radius inversely proportional to their distances.
Only the sources within $100$ Mpc are shown with bold lines.
Throughout the paper, we show the results only for this specific
source scenario.
We have checked that the results do not depend very much on the
realization of the source selection, unless extremely nearby sources
($< 30 {\rm Mpc}$) are accidentally selected.
We note that the current AGASA data set include $49$ events with
energies of $4\times 10^{19} - 10^{20}$ eV in the range of
$-10^{\circ} \le \delta \le 80^{\circ}$.
This event number corresponds to $100$ events in
figure~\ref{event}, where we do not restrict the arrival directions of
UHECRs in any range of $\delta$.

\begin{figure*}
\begin{center}
\epsscale{1.2} 
\plotone{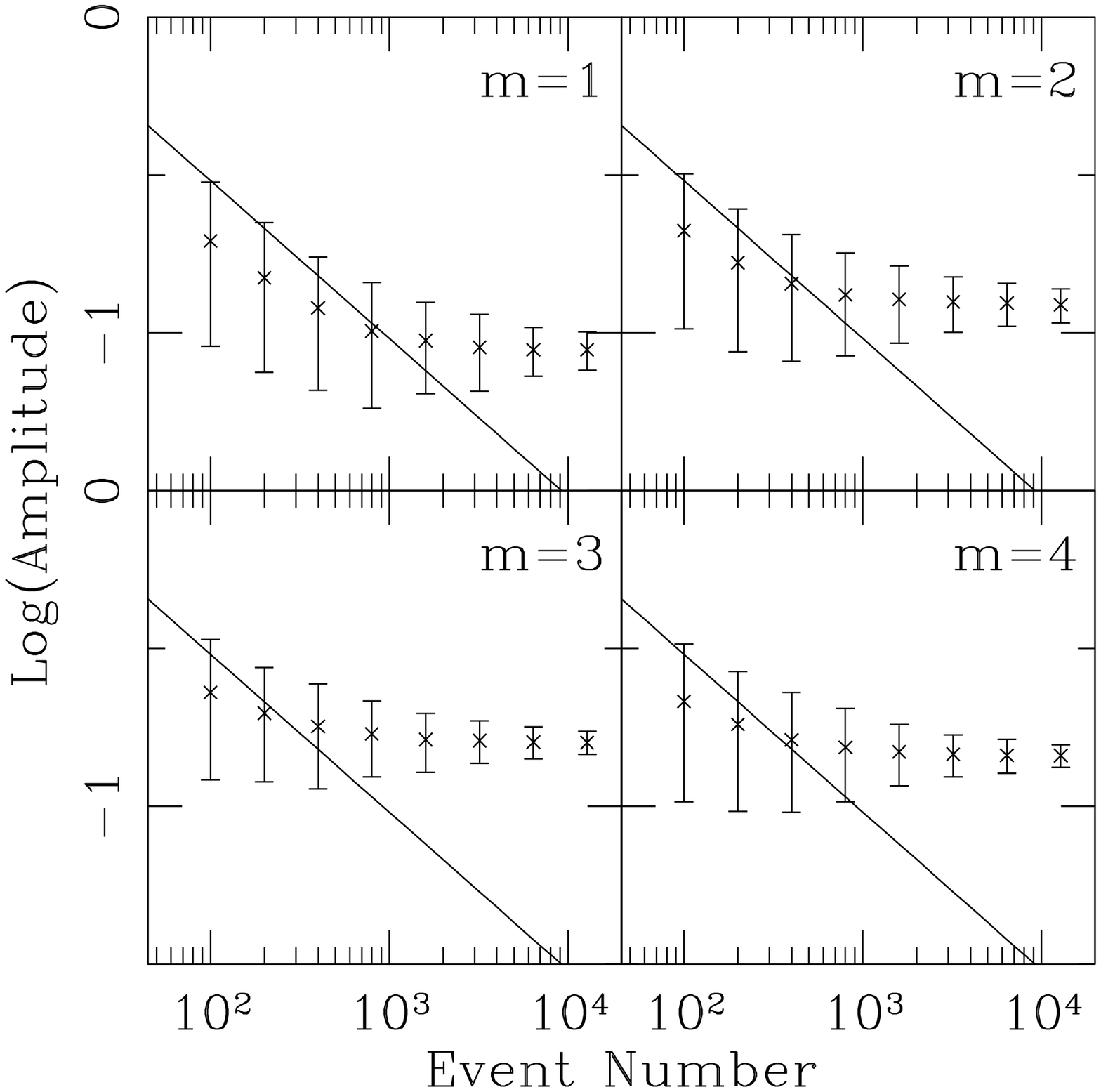} 
\caption{
Harmonic amplitude predicted by a source model of
Figure~\ref{event} as a function of the total number of events.
The errorbars represent the statistical fluctuations due to the finite
number of the simulated events.
The region below the solid line is expected from the
statistical fluctuation of isotropic source distribution with the
chance probability larger than 10$\%$.
\label{amp}}
\end{center}
\end{figure*}

A visual inspection of figure~\ref{event} reveals no significant large
scale anisotropy.
We show the harmonic amplitude as a function of the event
number for m$=1-4$ in figure~\ref{amp}.
We plot the average over all trial of the event realizations from the
calculated probability distribution with the statistical error.
In order to obtain the average and the variance, we dial the simulated
sets of events $20-1000$ times depending on the total event number.
The region below the solid line in this figure is expected from the
statistical fluctuation of isotropic source distribution with the
chance probability larger than 10$\%$.
Of course, this region become smaller with increasing the event
number.

At present, our source model predicts the harmonic amplitude
consistent with the isotropic source distribution.
However, future experiments will separate our model prediction from
the isotropic source at the confidence level of $90 \%$ with the event
number of the order $\sim 10^3$.
It is also found that the amplitude for $m=1$ is smaller than that for
another values of $m$.
The amplitude for larger $m$ quantifies anisotropy on smaller scale.
Therefore, this dependence on $m$ reflects the fact that the arrival
distribution shown in figure~\ref{event} becomes to reveal the
distribution of their sources with increasing the event number, in the
manner that the event clusterings occur at the directions of the
nearby ($< 100 \; {\rm Mpc}$) sources.
This feature of the UHECR arrival distribution is discussed below in
detail.

\begin{figure*}
\begin{center}
\epsscale{1.2} 
\plotone{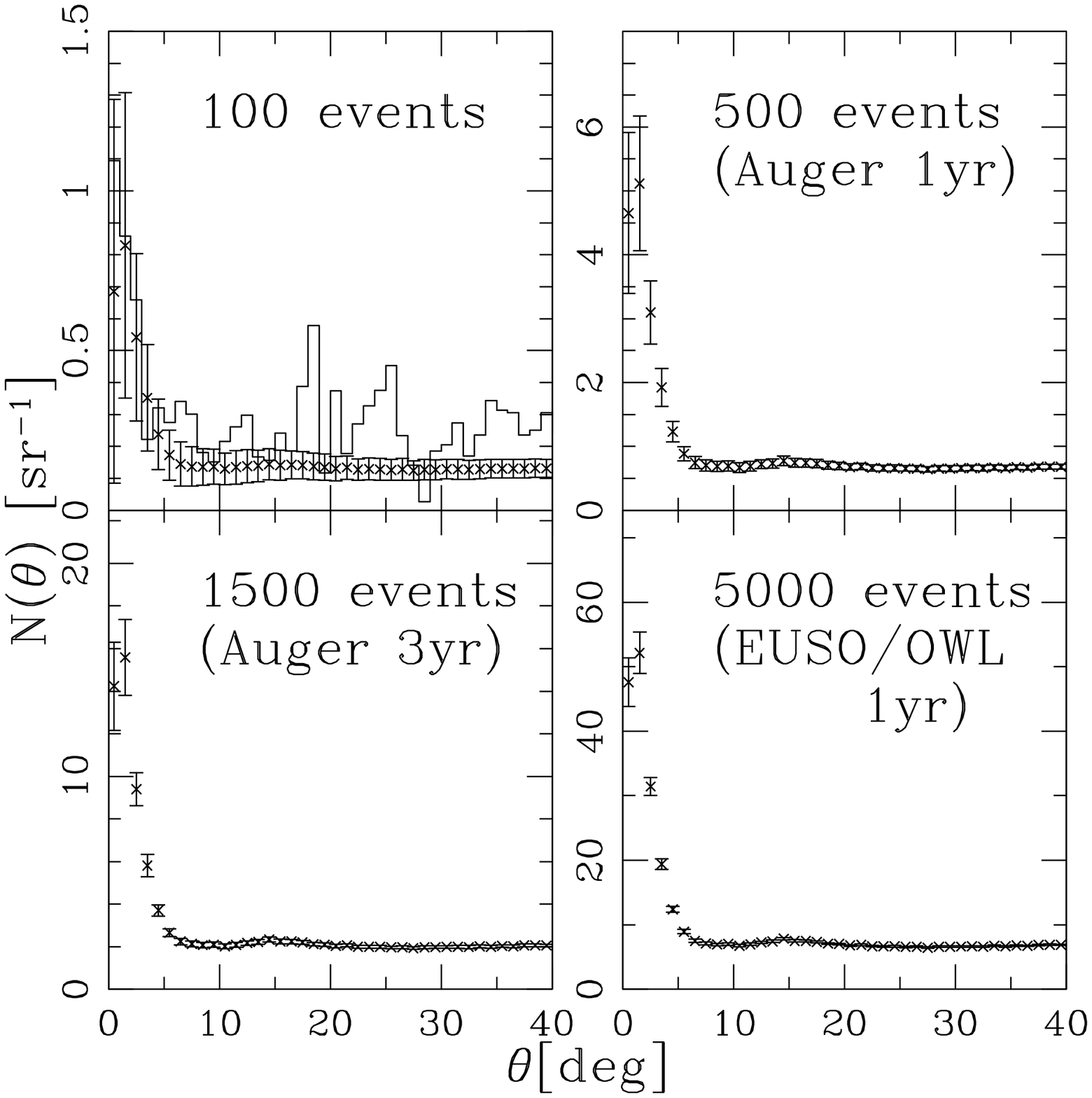} 
\caption{
Two point correlation functions predicted by a source model of
Figure~\ref{event}.
The errorbars represent the statistical fluctuations due to the finite
number of the simulated events.
For the event number of $100$, we also show the two point correlation
function on the AGASA 49 events in the energy range of $4\times
10^{19} - 10^{20}$ eV, multiplied by factor 2 in order to compensate the
difference of the range of $\delta$ between the observation and the
numerical calculation.
\label{2p}}
\end{center}
\end{figure*}

In this source scenario, the small scale anisotropy observed by the
AGASA is also reproduced, as is evident from figure~\ref{2p}, where
the two point correlation functions of the simulated events are shown.
The errorbars represent the statistical fluctuations due to the finite
number of the simulated events.
For the event number of $100$, we also show the two point correlation
function on the AGASA 49 events in the energy range of $4\times
10^{19} - 10^{20}$ eV, multiplied by factor 2 in order to compensate the
difference of the range of $\delta$ between the observation and the
numerical calculation.
Slight shrinkage of $N({\theta})$ at the smallest angle bin is due to
manner of dividing the sphere into concentric bins when taking the
data of numerical simulations of UHECR propagation.

Since the small scale anisotropy is due to the point-like nature of
UHECR sources, the angle scale at which there is strong correlation
between the events corresponds to the deflection angle of UHECRs in
propagation in the EGMF from sources to the earth.
Figure~\ref{2p} demonstrates that, for the amount of data expected
with next generation experiments, the statistical uncertainty will
considerably decrease.
This clarify the angle scale at which the events have strong
correlation with each other.
Accordingly, we will be able to determine the strength of the universal
EGMF from the two point correlation function of observed UHECRs with
the sufficient amount of data.

\vspace{0.1cm}
\centerline{{\vbox{\epsfxsize=8.0cm\epsfbox{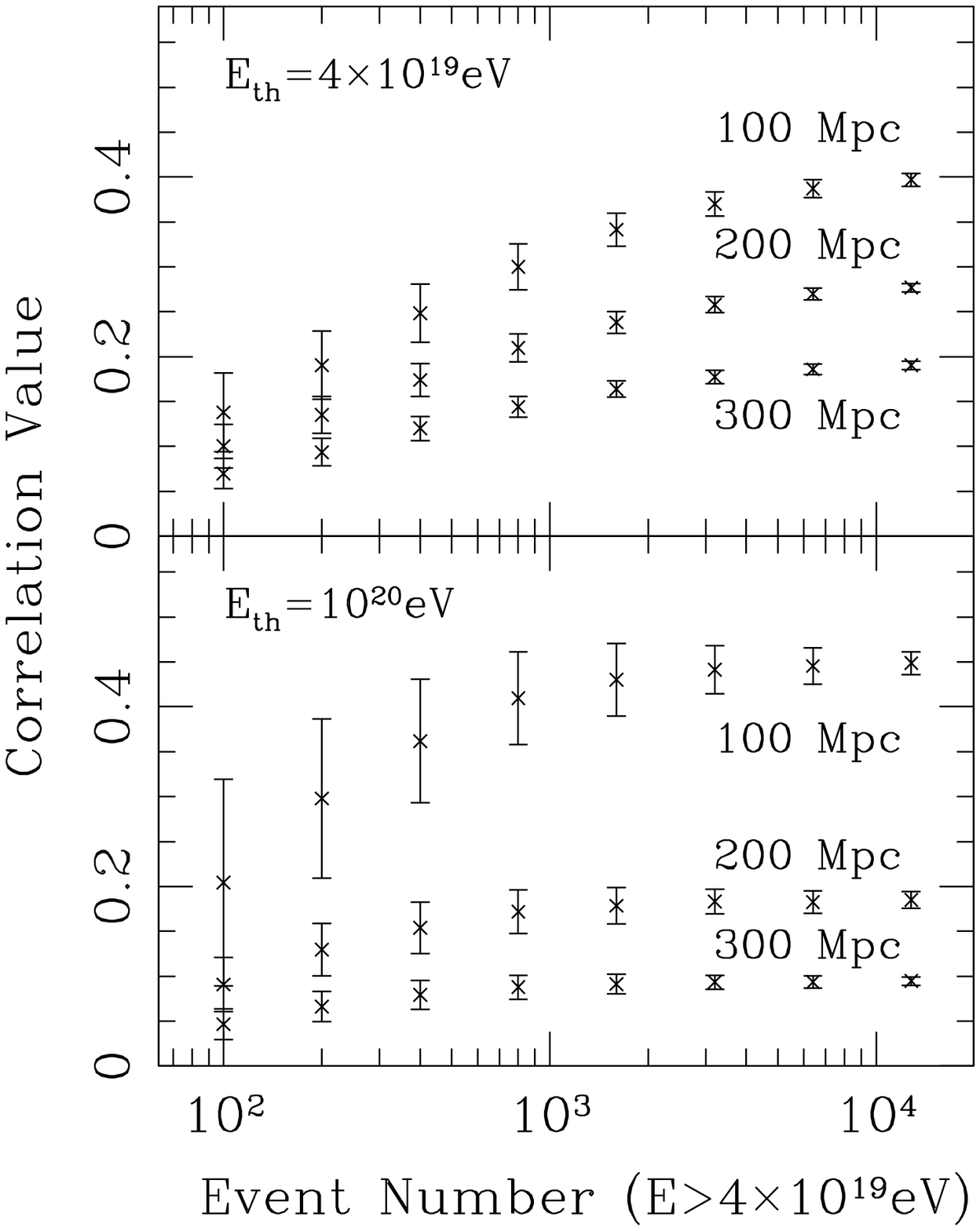}}}}
\figcaption{
Correlation value between the simulated events above $4\times
10^{19}$, $10^{20}$ eV and the source distribution of
Figure~\ref{event}.
The results are shown for the sources within 100, 200, and 300 Mpc
from us.
The errorbars represent the statistical fluctuations due to the finite
number of the simulated events.
\label{cor}}
\vspace{0.5cm}

As mentioned above, our source model predicts the statistically
significant small scale anisotropy, which correlate with the sources
located within $100$ Mpc, with a large number of data
(see, figure~\ref{event}).
What is deduced from this feature on the origin of UHECRs?
To begin with, we quantitatively examine the relation between the
event distribution and the source distribution.
The upper panel of figure~\ref{cor} shows the correlation value
defined in section~\ref{statistics} between the distribution of events
above $4 \times 10^{19}$ eV and the distribution of sources within
100, 200, and 300 Mpc as a function of the event number.
The results for the events above $10^{20}$ eV are discussed in the
next subsection.
The errorbars are same ones as that in Figure~\ref{amp}.

Clearly visible in this figure is that the correlation of the event
distribution with the sources are strongest for the sources within
$100$ Mpc.
This strong correlation is due to the event sets which cluster in the
direction of the sources within 100 Mpc (See Figure~\ref{event}).
The number of clustered events fluctuates every realizations, and this
number is a critical factor for the correlation with the sources
within 100 Mpc.
On the other hand, this number do not affect the correlation with the
sources within larger distances very much, because there are a number
of sources in this case.
Therefore, the statistical error is smaller for the correlation with
the sources at larger distances.
When the event number is close to the order of several $\times 10^3$,
the correlation values begin to converge, and the final values can be
estimated.
We emphasize that the expected event rate by the Auger observation is
$\sim 500$ per year.
After several years from the operation of the Auger, we would be able
to know the source distribution within $\sim$ 100 Mpc.

\begin{figure*}
\begin{center}
\epsscale{1.2}
\plotone{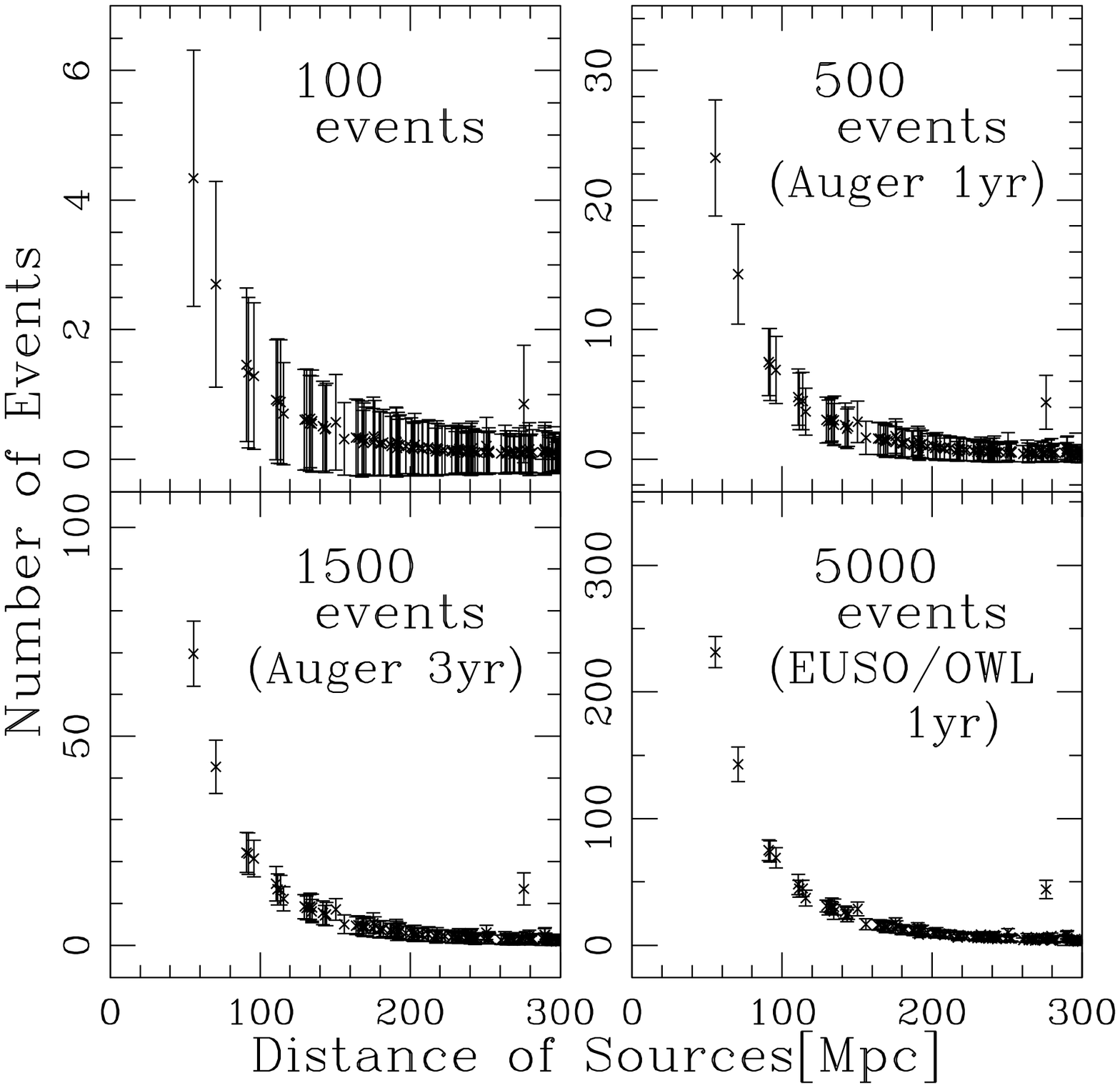} 
\caption{
Number of events above $4\times 10^{19}$ eV within $2.5^{\circ}$ from
the directions of UHECR sources of Figure~\ref{event} as a function of
the source distance.
The error bars represent the statistical error due to the finite
number of simulated events.
\label{source_clus}}
\end{center}
\end{figure*}

Here we note that the UHECR sources outside $107(=80 h^{-1})$ Mpc in
our model are randomly distributed, as mentioned in
section~\ref{source}.
We should compare the results of the correlation function between the
simulated events and the actual galaxy distribution with that between
the observed events and the actual galaxy distribution.
For the reason of the flux limit, we can not do so using the ORS
galaxy sample.
However, there is the galaxy survey with a limiting magnitude much
deeper than that of the ORS, Sloan Digital Sky Survey
\citep*[SDSS,][]{stoughton02}.
We conducted a study of the galaxy number density based on a
comparison of the observed number counts between the ORS and the SDSS
Early Data Release \citep*{yoshiguchi02b}.
As data obtained by the SDSS accumulate, we will be able to know the
actual galaxy distribution of much larger volume, and make
precise comparisons between numerical calculations and the
observations.

In order to further investigate the relation between the number of the
clustered events and the source distance, we calculate the number of
events above $4\times 10^{19}$ eV within $2.5^{\circ}$ from the
direction of each source of Figure~\ref{event}.
The angle $2.5^{\circ}$ roughly corresponds to both the observational
error of arrival directions and the deflection angle of UHECR when
propagating in the EGMF ($B=1 {\rm nG}$) over $\sim 100$ Mpc.
The result is shown in Figure~\ref{source_clus}.
The solid angle of the observer viewed from distant sources are
inversely proportional to the square of the distance.
Thus, contribution to cosmic-ray flux from a source closer to our
galaxy is larger than that from a distant source.
This is reflected in Figure~\ref{source_clus}, where the event number
at the direction of nearby sources are larger than that of distant
sources.
A source at $\sim$ 280 Mpc which have larger number of event in its
direction happens to be located at the direction of a closer source.

We again note that the event number $100$ corresponds to the observed
one by the AGASA experiment.
From figure~\ref{source_clus}, there must be a source at the direction
of triplet event sets within 100 Mpc from us at $2 \sigma$ confidence
level.
This implies that the triplet observed by the AGASA would originate
from sources within $100$ Mpc.
Indeed, \cite{smi02} show that there are merger galaxies
Arp 299 (NGC 3690 $+$ IC 694) at the direction of the AGASA triplet at
$\sim 42$ Mpc.
Considering the analysis presented here, Arp 299 is the best candidate
of the UHECR source.

Increasing the event number decrease the statistical uncertainty as is
evident from figure~\ref{source_clus}, and thus
the relation between the source distance and the number of clustered
events in its direction becomes clear.
As data accumulate by future experiment, we can know the distance of
the source which contributes to a clustered event set by using this
relation.
Performing this procedure for all the event clusterings, we would be
able to determine the distribution of UHECR sources within about 100
Mpc.

\subsection{Arrival Distribution of UHECRs above $10^{20}$ eV}
\label{arrival20}

In this subsection, we present the results of the arrival distribution
of UHECRs above $10^{20}$ eV.
It is noted that our source model predicts the cosmic-ray spectrum
with the GZK cutoff, unless other components are introduced at this
energy range.
Accordingly, the features of arrival distributions that we present
here are expected to be observed by future experiments if the current
HiRes spectrum is correct.
If the AGASA spectrum is correct, UHECRs of top-down origin may
dominate the cosmic-ray flux at this energy range.
However, we may be able to extract UHECRs of bottom-up origin from
ones of top-down origin by using the feature of arrival distribution
of bottom-up UHECRs, as discussed in the final section.

\begin{figure*}
\begin{center}
\epsscale{1.4} 
\plotone{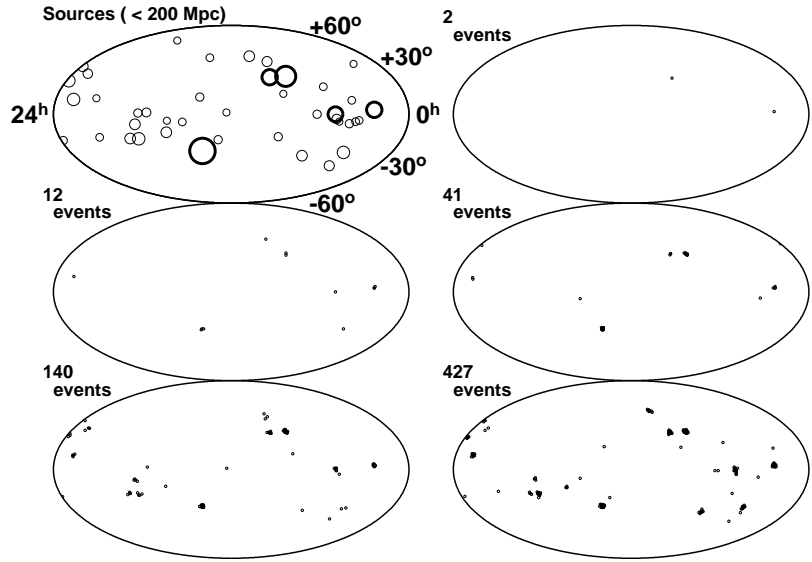} 
\caption{
Same as figure~\ref{event}, but only the events with energies above
$10^{20}$ eV are shown.
\label{event20}}
\end{center}
\end{figure*}

\begin{figure*}
\begin{center}
\epsscale{1.3} 
\plotone{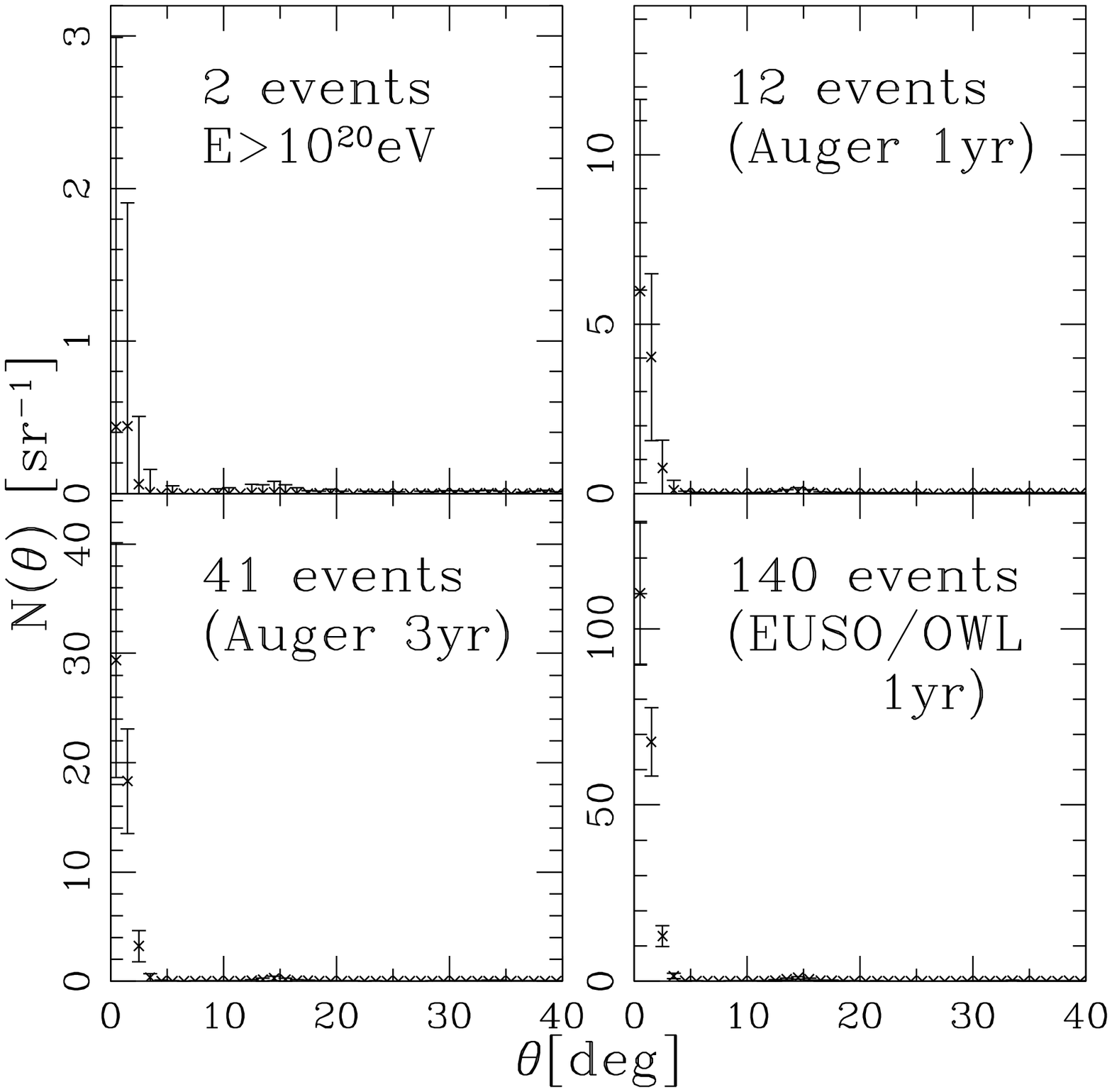} 
\caption{
Same as figure~\ref{2p}, but only for UHECRs above $10^{20}$ eV.
\label{2p_er5}}
\end{center}
\end{figure*}

In figure~\ref{event20}, we show the arrival distribution of UHECRs
above $10^{20}$ eV predicted by the source model of
figure~\ref{event}.
This figure is same as figure~\ref{event}, but only the events with
energies above $10^{20}$ eV are shown.
Compared to figure~\ref{event}, we find that the arrival directions
are concentrated in the few directions.
This can be also seen from figure~\ref{2p_er5}, which is the same
figure as figure~\ref{2p}, but for only UHECRs above $10^{20}$ eV.
At $\theta > 4^{\circ}$, the values of $N(\theta )$ are almost equal
to $0$.
This is because the sources of UHECRs above $10^{20}$ eV
must be located in a limited volume of radius at most $\sim 100$ Mpc
because of the pion production.
Sources at larger distances can not contribute to the cosmic-ray
spectrum above this energy.

This feature of the correlation between the events and the sources can
be also understood from the lower panel of figure~\ref{cor}.
Note that the horizontal axis of this figure is the event number above
$4\times 10^{19}$ eV.
Since the number of events above $10^{20}$ eV is very small, the
statistical error is larger for $E_{\rm th}=10^{20}$ eV than
$E_{\rm th}=4\times 10^{19}$ eV.
From this figure, it is clear that
the correlation with the sources within $100$ Mpc is stronger for
$E_{\rm th}=10^{20}$ eV than that for $E_{\rm th}=4\times 10^{19}$ eV.
On the other hand, this dependence becomes opposite in the case of the
sources within $200$ and $300$ Mpc, because UHECRs below $10^{20}$ eV
can come from a much larger volume ($\sim 1$ Gpc$^3$).

We also calculated the harmonic amplitude and the number of events
within $2.5^{\circ}$ from the directions of the sources for UHECRs
above $10^{20}$ eV.
The volume from which UHECRs at this energies can originate is much
smaller than that for $E<10^{20}$ eV.
Although the dependence of the harmonic amplitude on random
selection of UHECR sources from the ORS sample is relatively large for
this reason, significant anisotropy on large angle scale may be
observed when future experiments detect about $\sim {\rm several}
\times 10$ cosmic rays above $10^{20}$ eV.
The dependence of the number of events above $10^{20}$ eV within
$2.5^{\circ}$ from the directions of the sources on their distances is
almost same as that for above $4 \times 10^{19}$ eV.

\section{SUMMARY AND DISCUSSION} \label{summary}

In this paper, we predicted the arrival distribution of UHECRs above
$4 \times 10^{19}$ eV with the total number of events expected by next
generation experiments in the next few years.
We performed event simulations using the ORS galaxy sample to
construct a source model of UHECRs, which can explain the current
AGASA observation below $10^{20}$ eV (Paper I).
It is noted that our prediction is not the exact arrival directions of
each UHECR but the statistical features of the arrival distribution,
because there are degrees of freedom of randomly selecting the UHECR
sources from the ORS sample and this sample does not contain galaxies
outside $107(=80 h^{-1})$ Mpc.
However, we will be able to know the actual galaxy distribution of
much larger volume by using the SDSS galaxy sample
\citep*{stoughton02} with increasing amount of data.

At first, we calculated the harmonic amplitude and the two point
correlation function for the simulated event sets.
We found that significant anisotropy on large angle scale will be
observed when $\sim 10^3$ cosmic rays above $4 \times 10^{19}$ eV are
detected by future experiments.
The Auger array will detect cosmic rays with this event number in
a few years after its operation.
The local enhancement model of UHECR sources \citep*{isola02,sigl03}
predicts large-scale anisotropy of UHECR arrival distribution that
reflects the spatial structure of the LSC, which can not be seen from
our model prediction (see figure~\ref{event}).
Thus we will be able to determine which model is favored from the
observations in a few years.
The statistics of the two point correlation function will also
increase, and the angle scale at which there is strong
correlation between the events corresponds to deflection angle of UHECR
in propagating in the EGMF.
Thus, it is expected that we will be able to know the strength of the
EGMF, using the two point correlation function for observed arrival
distribution of UHECRs, if our source and magnetic field model is
supported by the future experiments.

Next, we investigated the relation between the number of clustered
events and the distance of the source at their direction.
We found that the C2 triplet observed by the AGASA
\citep*{hayashida00} may originate from the source within 100 Mpc from
us.
Indeed, \cite{smi02} show that there are merger galaxies
Arp 299 (NGC 3690 $+$ IC 694) at the direction of the AGASA triplet at
$\sim 42$ Mpc.
Considering the analysis presented here, Arp 299 is the best candidate
of the UHECR source.

When the event number increases, the statistical uncertainty decreases
as is evident from figure~\ref{source_clus}, and thus
the relation between the source distance and the number of clustered
events in its direction becomes clear.
Using this relation, we will be able to determine the distribution of
UHECR sources within about 100 Mpc.

Identification of the sources of UHECRs is extremely important.
At first, this will provide some kinds of information about poorly
known parameters which influence the propagation of UHECRs, such as
extragalactic and galactic magnetic field, chemical composition of
observed cosmic rays.
Furthermore, this will give an invaluable information on mechanisms and
physical conditions which lead to acceleration of cosmic rays to
energies of order $10^{20}$ eV.
In particular, we showed that there was the strong correlation between
the arrival distribution of UHECR above $10^{20}$ eV and the source
distribution within $100$ Mpc.
If the cosmic-ray spectrum measured by the HiRes experiment is
correct, the UHECR arrival distribution similar to
figure~\ref{event20} will be observed by future experiments.
We will be able to know the maximum energies achieved by
cosmic rays in each identified object.
If the AGASA spectrum is correct, and cosmic-ray flux is
dominated by the component of top-down origin.
However, the number density of the supermassive particles, whose decay
product can be observed UHECRs above $10^{20}$ eV, is estimated as
$10^{36}$ Mpc$^{-3}$ in our galactic halo in order to explain the
observed flux, as we discussed in Paper I.
In this case, there would be no small scale anisotropy of
arrival distribution of UHECRs above $10^{20}$ eV which are generated
by the top-down mechanisms.
On the other hand, the arrival directions of bottom-up UHECRs
are strongly concentrated in the few directions, as presented in the
previous section.
We may be able to extract UHECRs of bottom-up origin from
ones of top-down origin, and obtain an information on the maximum
energies of cosmic rays.

In Paper I, we showed that the number density of UHECR sources may be
$\sim 10^{-6}$ Mpc$^{-3}$ in order to explain the UHECR arrival
distribution observed by the AGASA experiments.
Nevertheless, we could not know which of the astrophysical objects
mainly contributes to the observed cosmic-ray flux.
However, with the method to identify the sources of UHECR developed in
this paper, we will reveal their origin and obtain an useful
information on acceleration mechanism to the highest energy.



\acknowledgments
This research was supported in part by Giants-in-Aid for Scientific
Research provided by the Ministry of Education, Science and Culture
of Japan through Research Grant No.S14102004 and No.14079202.



\end{document}